\documentclass[english,twocolumn,showpacs,preprintnumbers,amsmath,amssymb]{revtex4}
\usepackage[T1]{fontenc}
\usepackage[latin9]{inputenc}
\usepackage{amsmath}
\usepackage{graphicx}

\makeatletter
\@ifundefined{textcolor}{}
{%
 \definecolor{BLACK}{gray}{0}
 \definecolor{WHITE}{gray}{1}
 \definecolor{RED}{rgb}{1,0,0}
 \definecolor{GREEN}{rgb}{0,1,0}
 \definecolor{BLUE}{rgb}{0,0,1}
 \definecolor{CYAN}{cmyk}{1,0,0,0}
 \definecolor{MAGENTA}{cmyk}{0,1,0,0}
 \definecolor{YELLOW}{cmyk}{0,0,1,0}
 }



\usepackage{babel}

\begin{document}

\title{Porter-Thomas distribution in unstable many-body systems }

\author{Alexander Volya}

\affiliation{Department of Physics, Florida State University, Tallahassee, FL
32306-4350, USA}

\date{\today}
\begin{abstract}
We use the continuum shell model approach to explore the resonance
width distribution in unstable many-body systems. The single-particle
nature of a decay, the few-body character of the interaction Hamiltonian,
and collectivity that emerges in non-stationary systems due to the
coupling to the continuum of reaction states are discussed. Correlations
between structures of the parent and daughter nuclear systems in the
common Fock space are found to result in deviations of decay width
statistics from the Porter-Thomas distribution. 
\end{abstract}



\maketitle
The Porter-Thomas distribution (PTD) \citep{Porter:1956} of transition
strengths is a central aspect of complex systems. This statistical
law was noted by many authors \citep{Brody:1981} to be valid more
generally than other predictions of the Random Matrix Theory from
which it originates. The PTD emerges under the assumption that the
relative orientation of the two states involved in the overlap describing
a transition covers the $\Omega$-dimensional sphere in the Hilbert
space uniformly. While this is true by definition for the Gaussian
Orthogonal Ensemble (GOE), the validity of the PTD extends much farther
as it constitutes the central limit theorem (CLT). Being a sum of
a large number of uncorrelated components, the transitional amplitude
is indeed expected to have a Gaussian distribution. There is a large
volume of work on this subject; see reviews \citep{Brody:1981,Zelevinsky:2004,Papenbrock:2007,Mitchell:2010,RevModPhys.81.539}
and references therein. Generally, there is a consensus among authors
that while the specifics of an ensemble and the physics of transitions
do matter for certain observables, any deviations from the PTD are
quickly defeated by even small stochastic components due to the robust
nature of the CLT; see for example Ref. \citep{Grimes:1983}. Any
claims to the contrary, either experimental \citep{PhysRevLett.105.072502}
or theoretical \citep{Celardo:2010ru,Whitehead:1978}, have always
ignited debates and discussions \citep{Reich:2010}. 

Here we will not consider the data handling procedures, which on many
occasions, have been deemed to be the most likely reasons for the
deviations observed experimentally \citep{RevModPhys.81.539}. Instead,
we focus on the possible reasons for high correlation between the
transitioning states from the theoretical perspective. We analyze
the feasible scenarios with the help of the continuum shell model
approach \citep{Volya:2009,Volya:2003PRC} which is one of the most
equipped methods to address the structure-reaction physics of interest
microscopically. In a unified picture we review the superradiance
(SR) effects \citep{Celardo:2010ru,Muller:1988}, the wave-function
localization effects in the two-, three-, and four-body ensembles
\citep{Kaplan:2000}, the role of rotational symmetry, and other parent-daughter
structural correlations that emerge in a decay \citep{Grimes:1983}.

The dynamics of an unstable many-body system projected onto the intrinsic
space spanned by the bound (shell-model) states is generated by the
effective, energy-dependent Hamiltonian \citep{Feshbach:1991,Mahaux:1969,Volya:2009}
\begin{equation}
{\cal H}=H-i\sum_{c({\rm open})}{\wp}^{c}\,|c\rangle\langle c|.\label{eq:HAM-1}\end{equation}
Here $H$ is the Hermitian part that is identified with the traditional
shell model Hamiltonian. The second, imaginary term reflects the irreversible
decays into the continuum of states excluded by the Feshbach projection.
This factorized operator contains the kinematic penetrability factor
${\wp}^{c}$ and the set of channel vectors $|c\rangle.$ For simplicity
we omit the angular momentum, isospin, and other labelings; detailed
notations are found in Ref. \citep{Volya:2009}. The problem is non-stationary;
the Hamiltonian \eqref{eq:HAM-1} is understood as a component of
the propagator and is dependent on the scattering energy. The Hermitian
component includes the coupling to the continuum of reaction states
via virtual excitations; the penetrability also depends on energy
through the kinematics of the decay process. Away from thresholds
the energy dependence is smooth, and its exact form mainly pertains
questions of the experimental data analysis. We ignore this dependence
here, and further assume that the penetrability is the same for all
channels. The eigenvalues of the effective Hamiltonian \eqref{eq:HAM-1}
are complex, ${\cal E}=E-i\Gamma/2,$ and represent the poles of the
scattering matrix in the complex energy plane. These complex energies
are associated with resonances and their widths. 

Let us first consider weak decays, for which the imaginary component
in \eqref{eq:HAM-1} can be treated perturbatively. In this case the
shell model eigenstate $|I\rangle$ defined by $H|I\rangle=E_{I}|I\rangle$
is not modified by the decay instability, and the corresponding decay
width is \[
\Gamma_{I}=2{\wp}\,\gamma_{I},\quad{\rm where}\quad\gamma_{I}=\sum_{c({\rm open})}\left|\langle I|c\rangle\right|^{2}\]
is the reduced width. The PTD of reduced widths \begin{equation}
P_{\nu}(\gamma)=\frac{1}{\gamma}\left(\frac{\nu\gamma}{2\overline{\gamma}}\right)^{\nu/2}\frac{1}{\Gamma(\nu/2)}\,\exp\left(-\frac{\nu\gamma}{2\overline{\gamma}}\right)\label{eq:PTD}\end{equation}
emerges under the uniform Hilbert space coverage assumption for $|I\rangle.$
Here $\nu$ is the dimension spanned by the channel vectors, and $\overline{\gamma}$
is the average reduced width. For the orthogonal and normalized channels
$\overline{\gamma}=\nu/\Omega$. 

In this work we examine situations with only one open channel. The
strength of the continuum coupling is defined via the average decay
width relative to the level spacing. Here we express this coupling
using the parameter $\kappa={\wp}/\lambda,$ where $\lambda^{2}=\Omega^{-1}{\rm Tr}\left(H^{2}\right)$
is the variance of the density of states distribution of $H$. 

\begin{figure}
\includegraphics[width=7cm]{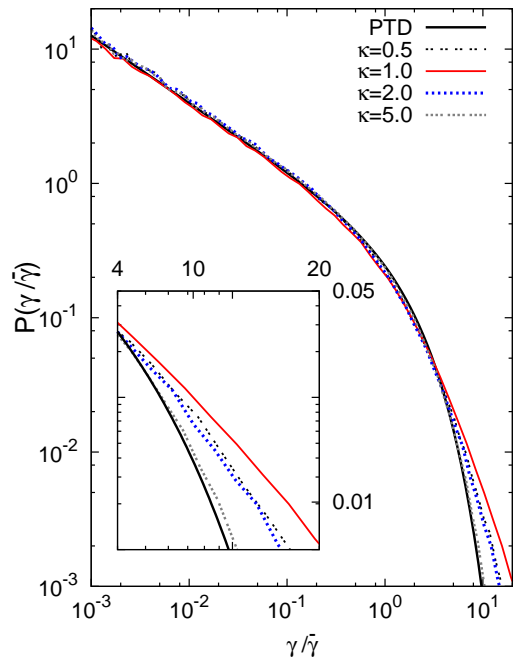}\caption{(Color online) The normalized distribution of probabilities of the
reduced widths for the GOE+SR ensemble with $\Omega=10^{4}.$ The
curves for different continuum couplings $\kappa=0.5,$ 1, 2, and
5 are compared with the PTD. The corresponding average widths $\Omega\overline{\gamma}=0.76,$
0.40, 0.20, and 0.05, are computed after the very broad state(s) are
removed from consideration. For $\kappa>2$ this amounts to exclusion
of a single superradiant state in which case $\Omega\overline{\gamma}=\kappa^{-2}.$
With the exception of very broad states, there is perfect agreement
with PTD, the differences being magnified in the inset where the region
of $\gamma>4\overline{\gamma}$ is shown. The largest deviation is
observed for the transitional coupling strength $\kappa=1$. For $\kappa=5$
the distribution is already indistinguishable from the PTD.\label{fig:sr}}

\end{figure}

In Fig. \ref{fig:sr} we consider an example of GOE+SR, where $H$
in (\ref{eq:HAM-1}) is represented by the GOE. Here the PTD is reproduced
numerically in the limit ${\wp}\rightarrow0.$ The density of states
in this limit has a semicircular distribution bound by a radius $2\lambda.$
The imaginary component in (\ref{eq:HAM-1}) is factorized, which
is pertinent to the unitarity of the scattering matrix. The non-Hermitian
component, when large, gives rise to the collectivity often referred
to as superradiance (SR). A similar collectivity due to the factorized
Hermitian interaction describes giant resonances. The resulting deformed
random ensembles are discussed in Refs. \citep{RefWorks:1468,Brody:1981,Sokolov:1988}.
As coupling to the continuum increases, and the average decay width
relative to the level spacing $\kappa$ becomes large, the resonances
start to overlap, thus reorienting the intrinsic structure. This could
be hypothesized to result in the PTD being violated \citep{Celardo:2010ru}.
An in-depth examination, however, shows that the SR mechanism alone
is unlikely to cause a significant change to the PTD. Indeed, for
$\kappa\ll1,$ the PTD simply follows from the definition of GOE.
For $\kappa\gg1,$ in full mathematical equivalence to the deformed
ensembles, the Hilbert space is separated into the SR channel space,
which is one-dimensional here with a single eigenstate ${\cal E}=-i{\wp},$
and the orthogonal statistical (compound resonance) space of dimension
$\Omega-1$. Because $H$ is orthogonally invariant the reduced-space
dynamics is represented by the GOE. With the perturbation theory built
in this limit one finds that the reduced widths for the compound resonances
follow the PTD with $\overline{\gamma}=1/(\Omega\kappa^{2}).$ For
large $\Omega$ the single SR state with a reduced width $\gamma_{SR}=1-\kappa^{-2}$
has no effect on the PTD. As seen in Fig. \ref{fig:sr}, numerical
study confirms the PTD for both small and large values of $\kappa.$
Moreover, the slight deviation from the PTD for couplings $\kappa$
between around 0.4 and 2 is due to a small fraction of very broad
states with $\gamma>10\overline{\gamma}$. This localized effect is
shown in the inset of Fig. \ref{fig:sr}. This subset of an exceptionally
broad states is difficult to identify experimentally. Investigations
\citep{RefWorks:1468,Brody:1981,Sokolov:1988} of deformed ensembles
further confirmed a good agreement of the decay width distribution
with the PTD and of the density of states distribution with the semicircular
shape. 

Experience with the realistic nuclear structure and some theoretical
arguments \citep{Brody:1981,Zelevinsky:2004,Papenbrock:2007,Mitchell:2010,RevModPhys.81.539}
suggest that the effective Hamiltonian involves only few-nucleon interactions,
thus, the two-body random ensembles (TBRE) appear to be more appropriate.
Many features of these ensembles are different from those of GOE,
nevertheless the eigenvectors form a uniform coverage of the Hilbert
space. Numerical studies confirm that, in agreement with the CLT,
this leads to the PTD of transition strengths toward an uncorrelated
channel vector \citep{Grimes:1983}. This logic, however, does not
take into account the correlations that exist in the variable particle-number
Fock space. Strong parent-daughter correlations emerge due to the
microscopic physics of decay. Indeed, in the single-particle reaction
processes all nucleons, except for one, are spectators, and the decay
channels for the $N-$particle system $|c;N\rangle$ are built from
the $(N-1)$-particle eigenstates of the daughter nucleus $|F;N-1\rangle$
that follow from the same two-body Hamiltonian. Thus, $|c;N\rangle=\left\{ a^{\dagger}|F;N-1\rangle\right\} ,$
where $a^{\dagger}$ is a single-particle creation operator corresponding
to the decaying nucleon, and brackets \{...\} indicate normalization
to unity and appropriate symmetry coupling. The correlation between
eigenstates and channels leads to the violation of PTD. Single-particle
removal amplitudes are related to the independent-particle basis,
where departure from the PTD has been demonstrated in the past \citep{Whitehead:1978}. 

\begin{figure}
\includegraphics[width=7cm]{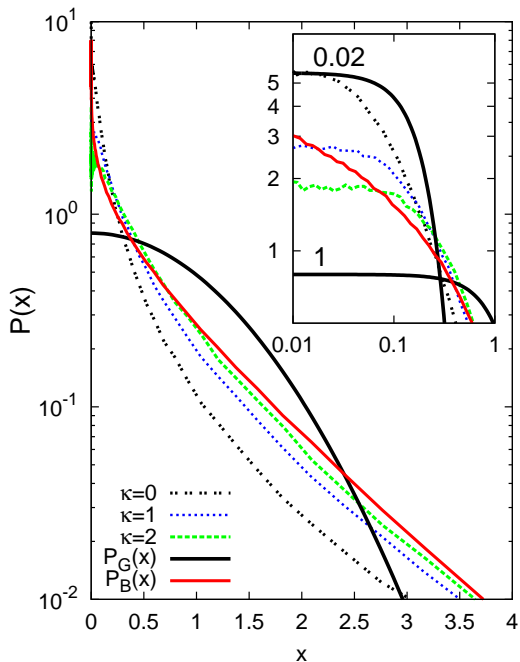}\caption{(Color online) The normalized distribution of probabilities of decay
amplitudes in the EGOE+SR ensemble. The decay of a 7-particle system
to the ground state of a 6-particle system is considered. There are
16 single-particle states, thus $\Omega=11440$. The curves for different
continuum couplings $\kappa=0,$ 1, and 2 are compared with the Gaussian
and Bessel distributions. The curves are normalized so that the average
widths do not include the single SR state. $\Omega\overline{\gamma}=1,\,0.19,$
and 0.06 for $\kappa=0,\,1$ and 2 respectively. The region of very
narrow widths is shown in the inset using a log-log scale. While all
observed distributions for very narrow states seem to approach a constant,
they are still not described by Gaussian distribution of different
variances (or $\overline{\gamma}$'s). The inset includes two Gaussian
curves with variances 1 and 0.02 as labeled.\label{fig:eax}}

\end{figure}

In Fig. \ref{fig:eax} the EGOE+SR ensemble is considered, where $H$
in Eq. (\ref{eq:HAM-1}) is given by an embedded two-body GOE (EGOE),
and by definition does not include any symmetries. The previous example
shows that it is important to draw attention to the statistics of
the narrow states. In this work we do not present our fits of the
observed distributions to the PTD (\ref{eq:PTD}) treating $\nu$
as a parameter. We found that attempts to do so could produce misleading
results. The effective $\nu$ is directly related to the normalization
$\overline{\gamma}$ which is disproportionately influenced by statistically
unimportant collective state(s). In an attempt to avoid any potential
misjudgements, here we show the distribution of absolute values of
amplitudes $x=\sqrt{\gamma/\overline{\gamma}}$. Then the Gaussian
distribution $P_{G}(x)=\sqrt{2/\pi}\exp\left(-x^{2}/2\right)$ that
corresponds to the PTD is easy to separate from a distribution provided
by the Bessel function $P_{B}(x)=(2/\pi)K_{0}(x).$ In contrast to
the CLT, the latter distribution emerges when the transitional overlap
is possible only due to a single component in the wave-function along
some direction in the Hilbert space given by a vector $|1\rangle$;
so that $\langle I|c\rangle=\langle I|1\rangle\langle1|c\rangle,$
where both $\langle I|1\rangle$ and $\langle1|c\rangle$ are distributed
normally (agree with PTD).%

From Fig. \ref{fig:eax} we find that none of the EGOE+SR results
follow the PTD. In contrast to the Gaussian curve $P_{G}(x)$ the
distributions have sharp peaks at low amplitudes and an extended exponential
tail. In the SR limit of large $\kappa$ the distributions appear
to approach the one given by the Bessel function $P_{B}(x)$. 

The rank of the force beyond the two-body interaction, and symmetries,
such as rotational, may have additional influences. To examine this
we consider a model where $N$ identical fermions occupy a single-$j$
level. This has been a popular model for exploring the properties
of TBRE \citep{Johnson:1998,Zelevinsky:2004,Papenbrock:2007}. The
resemblance of the low-lying spectra to those observed in realistic
nuclei is the most intriguing feature. For our demonstration we select
$j=19/2$ and discuss widths of the decay of many-body states in $9$-particle
systems. The final state is the ground state of the system with $8$
nucleons. All states, in both parent and daughter nuclei, are eigenstates
of the same Hamiltonian given by the $n-$body Random Ensemble ($n$-BRE)
\citep{Volya:2008}. In this work we restrict our consideration to
the two-, three-, and four-body forces, $n=2,3,$ and 4, respectively.
We select only those realizations where the daughter system has ground
state spin $F=0;$ and thus the channel spin is $I=19/2$. The fractions
of such realizations are 42\%, 64\%, and 83\% for $n=2,3,$ and 4
respectively. 

\begin{figure}
\includegraphics[width=7cm]{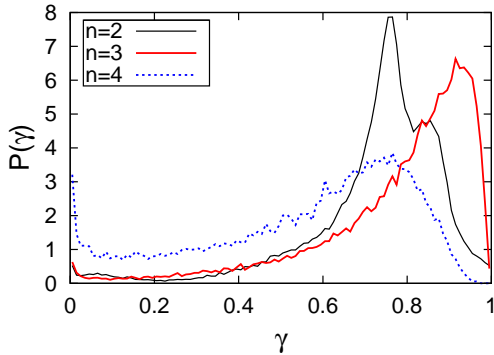}\caption{(Color online) Distribution of the reduced decay widths (spectroscopic
factors) for the ground state $N=9$ spin $I=19/2$ parent decaying
to the ground state $N=8$ spin $F=0$ daughter system. The $n$-BRE
of identical nucleons in $j=19/2$ level is considered. Three curves
correspond to two-, three-, and four-body ($n=2,3,$ and 4) random
ensembles with rotational symmetry.\label{fig:Identical-nucleons}}

\end{figure}

The resemblance between random ensembles with symmetries and realistic
nuclei extends to parentage relations. The low-lying states in the
odd-particle parent nucleus are predominantly of the single-particle
nature. If $F=0$ for the even-particle core then the ground state
of a system with an extra nucleon is likely to carry the single-particle
quantum numbers $j=I=19/2.$ This is indeed observed, and the corresponding
probabilities are 21\%, 47\%, and 37\% for $n=2,3,$ and 4 respectively.
The correlation between the parent and daughter ground states is demonstrated
in Fig. \ref{fig:Identical-nucleons} which shows the distribution
of reduced widths for the decay from ground state to ground state
when $F=0$ and $I=19/2$. Both parent and daughter systems have correlated
structures because they are eigenstates of the same Hamiltonian for
different number of particles. In the distribution of spectroscopic
factors this correlation is seen as a peak near $\gamma=1.$ As the
rank of interaction $n$ becomes higher, more remote configurations
can be admixed, which reduces the ground state to ground state transitional
collectivity. For totally uncorrelated systems the transitional strength
due to the CLT is expected to follow the PTD. 

\begin{figure}
\includegraphics[width=7cm]{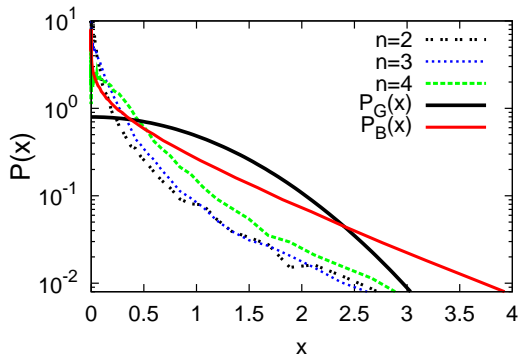}\caption{(Color online) Same model as in Fig. \ref{fig:Identical-nucleons}
The distribution of spectroscopic amplitudes for all 204 states with
spin $I=19/2$ in the parent $9$-nucleon system is shown in the decay
to the spin $F=0$ ground state of the daughter nucleus. The results
are compared with the Gaussian and Bessel distributions. The curves
do not change if states only in a certain energy region are considered.\label{fig:PT}}

\end{figure}

 Since most of the transitional strength is concentrated in a few
states at the low end of the spectrum, here the PTD is not generally
expected; however, there is also no agreement with this law for the
widths of highly excited compound resonances. In Fig. \ref{fig:PT}
the distribution of reduced decay amplitudes is shown for the two-,
three-, and four-body random ensembles with rotational symmetry. The
curves are quite close to each other and are similar to the EGOE result
($\kappa=0$ in Fig. \ref{fig:eax}). All findings indicate violation
of the PTD.

This study is motivated by the long-standing debate in relation to
the Porter-Thomas distribution and possibility of its violation \citep{Celardo:2010ru,Grimes:1983,PhysRevLett.105.072502,Reich:2010,Whitehead:1978}.
The PTD is a robust prediction justified by the central limit theorem;
it is easily confirmed for different random matrix ensembles. This,
however, is a purely structural approach that does not take into account
the microscopic physics of reactions. To address this we use a continuum
shell model approach where violations of the PTD may result from one
or a combination of the following: coherence in structure due to factorized
nature of the effective Hamiltonian that reflects unitarity of the
scattering matrix, the so called superradiance mechanism; parent-daughter
relation between decaying systems in the common Fock space; few-body
low rank interaction forces; and significant variations in the energy
dependence of the effective Hamiltonian. We examine all of these possibilities,
with the exception of the last, which is to be discussed elsewhere. 

In agreement with the studies of deformed ensembles we find that the
SR mechanism by itself does not lead to violations of the PTD. However,
the distribution, unambiguously different from PTD, is observed in
random ensembles with a particle decay and with few-body intrinsic
Hamiltonians. This remains true even when subsets of states are considered:
restricted by a certain energy region, or by some reasonable limits
on the value of the decay width itself. The latter allows for any
collective or unmeasurably broad states to be excluded. The parent-daughter
relation in the decay process appears to be central for this phenomenon,
the results being only slightly influenced by SR, additional rotational
symmetry or the rank of forces.

The author is thankful to Kirby Kemper for motivating discussions.
Support from the U. S. Department of Energy, grant DE-FG02-92ER40750
is acknowledged. The computing resources were provided by the Florida
State University shared High-Performance Computing facility.

\bibliographystyle{apsrev}
\bibliography{ptd}

\end{document}